\documentclass{article}

\usepackage{arxivstyle}

\usepackage{fontspec}
\usepackage{amsmath, amssymb, amsthm}
\usepackage{booktabs}
\usepackage{array}
\usepackage{multirow}
\usepackage{graphicx}
\usepackage{float}
\usepackage{caption}
\usepackage{placeins}
\usepackage[hidelinks]{hyperref}
\usepackage{url}
\usepackage{xcolor}
\definecolor{codegray}{gray}{0.95}
\newcommand{\code}[1]{\colorbox{codegray}{\texttt{\small #1}}}

\usepackage{xeCJK}
\setCJKmonofont{FandolFang-Regular.otf}
\xeCJKsetup{CJKecglue=}

\newcommand{\Tpos}{T^{+}}
\newcommand{\Tneg}{T^{-}}

\graphicspath{{./}{figs/}{figs/fig0_motivation/}{figs/fig1_hero/}{figs/fig2_gap/}{figs/fig3_main_gain/}{figs/fig4_e2e_heatmap/}{figs/fig5_field_contrib/}}

\title{Skills Know Their Neighbors: Cluster-Contrastive Capability Pages for Skill Retrieval}

\author{
  Zifei Wang$^{1,*}$,\ \ Wei Wen$^{2,*\dagger}$,\ \ Qiang Ji$^{1}$,\ \ Ruizhi Qiao$^{1,2}$ \\[2pt]
  $^{1}$Tencent IMA Product Center \quad $^{2}$Tencent Youtu Lab \\[2pt]
  \texttt{\{zifeiwang, jawnrwen, peasirji, ruizhiqiao\}@tencent.com}
}

\begin{document}
\maketitle
\lhead{}\chead{}\rhead{}
\pagestyle{plain}\thispagestyle{plain}
\renewcommand{\thefootnote}{\fnsymbol{footnote}}
\footnotetext[1]{Equal contribution.}
\footnotetext[2]{Corresponding author. Contact: \texttt{jawnrwen@tencent.com}.}
\renewcommand{\thefootnote}{\arabic{footnote}}
\setcounter{footnote}{0}

\begin{abstract}
As skill libraries grow, large language model agents must retrieve reusable skills from candidates that often share the same topic and vocabulary but implement different capabilities. Retrieval is limited not only by the scorer but also by the text being scored: a document may describe what a skill does without stating which similar requests should be routed elsewhere. We formalize a skill's capability as its \emph{executable region}, the set of queries it can solve, and view its document as a lossy observation of that region. This view exposes a document-imposed component of retrieval error that cannot be removed by improving the retriever alone. We therefore propose \emph{Capability Pages}, cluster-contrastive skill representations containing a positive trigger $\Tpos$, a negative boundary $\Tneg$, and a discriminative body $B$. An offline compiler compares neighboring skills to write these fields. At inference time, the index uses $\Tpos$, $B$, and the original document for candidate recall, while the router uses $\Tneg$ to reject confusable alternatives. On SRA-Bench, which contains 26{,}262 skills and 5{,}400 questions from six datasets, Capability Pages improve Recall@10 for all five tested retrievers, with a mean gain of $2.94$ points. Adding $\Tneg$ to candidate cards improves end-to-end task success by $3.62$ points on average across four executors and six datasets. A transfer evaluation on Chinese SSL-SkillDiscovery reaches $73.07\%$ MRR@50 using the same encoder across conditions. Capability Pages require no modification to the online models; they improve routing by rewriting the offline skill library.
\end{abstract}

\section{Introduction}

Reusable skills package task-specific instructions, constraints, tools, and execution logic for large language model (LLM) agents. When a library contains only a few skills, an agent can inspect all candidates or rely on explicit names. At larger scales, however, the system must first retrieve a small candidate set and then select an executable skill. This routing decision is a critical bottleneck: if the correct capability is absent from the candidate set or mistaken for a close neighbor, downstream reasoning has little opportunity to recover.

Existing work primarily improves the retrieval pipeline through stronger encoders, skill-specific training, or reranking. These approaches assume that the indexed skill text already exposes the evidence needed for discrimination. In real libraries, that assumption often fails. A per-flight lookup skill and a route-level flight-count skill may both mention flights, airports, and dates. Two mathematical skills may share nearly all formula-related terminology yet differ in a single decision condition. Such pairs are topically similar by design, yet only one may execute a given request.

\begin{figure}[H]
\begin{minipage}[t]{0.46\columnwidth}
\vspace{0pt}%
\centering
\includegraphics[width=\linewidth]{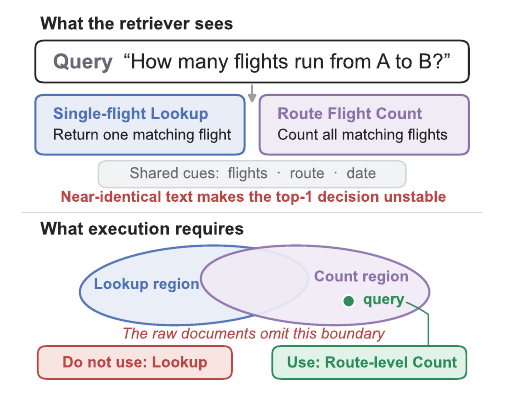}
\captionof{figure}{Similar skill text can hide different executable regions, making top-1 routing unstable.}
\label{fig:motivation}
\end{minipage}\hfill
\begin{minipage}[t]{0.50\columnwidth}
\vspace{0pt}%
The resulting error is not purely a model-capacity problem. A conventional skill document explains the supported procedure, but it rarely states the boundary between that procedure and neighboring capabilities. Consequently, two documents can be nearly indistinguishable to a retriever even when the corresponding skills solve different query sets. Scaling the encoder can improve how accurately it reads the available text, but it cannot recover distinctions that the text never expresses. Figure~\ref{fig:motivation} illustrates this mismatch: the documents share the same surface cues, but routing requires an explicit boundary between their supported query sets.\par\vspace{3pt}

We study this missing-information problem by separating a skill's capability from its document. Define a skill's capability as its \emph{executable region} $\mathcal{R}(s_i)$, the set of queries the skill can solve. The

\end{minipage}
\end{figure}

\vspace{-3pt}%
\noindent document $x_i$ is then a lossy observation of this region. This formulation yields a document-imposed lower bound: when two skills have indistinguishable documents but different executable regions, no scoring function of the raw documents can separate them on every query. Reducing this error requires changing the observation rather than only changing the scorer.

To this end, we introduce \emph{Capability Pages}, an offline representation compiled by comparing neighboring skills. Each page contains (i) a positive trigger $\Tpos$ describing requests that should select the skill, (ii) a negative boundary $\Tneg$ describing confusable requests that belong to a neighboring skill, and (iii) a discriminative body $B$ that states the skill's defining rule. The system exposes two views of the page: the retrieval index uses $\Tpos$, $B$, and the original document to improve candidate recall, whereas the router receives $\Tneg$ as an explicit exclusion condition. This separation avoids contaminating similarity embeddings with negative descriptions while preserving the value of $\Tneg$ for candidate-level comparison. Unlike concurrent work that retains WRITE/SKIP labels to train retrieve--rerank models under fixed skill documents~\cite{wang2026skillisnotdoc}, we leave the online scorer unchanged and change the indexed text.

Experiments evaluate both stages of the routing pipeline. On the 26{,}262-skill SRA-Bench, the index view improves Recall@10 for BM25, TF-IDF, BGE-M3, and two Qwen3-Embedding models, with a mean gain of $2.94$ points. In end-to-end evaluation, adding $\Tneg$ improves task success by $3.62$ points on average over 24 executor--dataset combinations. The gains also transfer to Chinese SSL-SkillDiscovery, where this representation reaches $73.07\%$ MRR@50 using the same encoder across conditions.

Our contributions are threefold:
\begin{itemize}
  \item \textbf{A text-side formulation of skill-retrieval error.} We model skills as executable regions and documents as lossy observations, derive a diagnostic decomposition of retrieval error, and state a clarifying lower bound for functionally different skills with indistinguishable documents.
  \item \textbf{Cluster-contrastive Capability Pages.} We propose an offline compiler that writes $\Tpos$, $\Tneg$, and $B$ from neighborhoods of confusable skills, together with a two-view design that uses positive evidence for retrieval and negative boundaries for routing.
  \item \textbf{Consistent gains across retrieval, execution, and transfer.} Capability Pages improve first-stage Recall@10 across five retrievers, end-to-end task success across four executors, and cross-distribution retrieval on Chinese SSL-SkillDiscovery without changing the online models.
\end{itemize}

\section{Related Work}

\paragraph{Skill retrieval and routing.}
Recent benchmarks frame skill selection as a distinct problem in agent pipelines. SRA-Bench, SkillsBench, SkillRet, R3-Skill, and SkillResolve-Bench evaluate retrieval or downstream skill use across different library scales~\citep{su2026srabench,li2026skillsbench,cho2026skillret,wang2026skillisnotdoc,ding2026skillresolvebench}. SkillRouter and SkillFlow directly target skill retrieval~\citep{zheng2026skillrouter,tagkopoulos2025skillflow}, while Skill-RAG and ToolOmni study routed retrieval strategies and tool selection in adjacent settings~\citep{wei2026skillrag,huang2026toolomni}. These methods match queries against fixed skill representations. Our work instead keeps the online retriever and router unchanged and improves the text they consume.

\paragraph{Document expansion and capability representation.}
Document-expansion methods such as doc2query and doc2query$--$ enrich retrieval text with likely queries, while HyDE constructs a hypothetical document from the query~\citep{nogueira2019doc2query,gospodinov2023doc2queryminus,gao2022hyde}. These approaches can improve positive coverage, but they process an item or query independently. Capability Pages instead define $\Tneg$ relative to neighboring executable regions, so it cannot be derived from one document alone. Structured skill representations such as SSL, Corpus2Skill, Graph-of-Skills, SkillDAG, and SkillSmith expose metadata, internal structure, or dependencies~\citep{liang2026ssl,sun2026corpus2skill,liu2026graphofskills,bai2026skilldag,xu2026skillsmith}. We focus on discriminative text for first-stage retrieval and candidate-level routing.

\paragraph{Skill execution and reuse.}
Agent frameworks study how selected skills support reasoning, execution, and reuse. ReAct interleaves reasoning with actions, while SkillX, Trace2Skill, and SkillWiki address skill construction or execution after selection~\citep{yao2023react,wang2026skillx,ni2026trace2skill,huang2026skillwiki}. Our method operates before execution, aiming to deliver a skill whose executable region contains the query without changing the executor.

\section{Problem Formulation}
\label{sec:formulation}

We distinguish the capability implemented by a skill from the document used to retrieve it. This distinction identifies which errors can be reduced by a better scorer and which require a better representation.

\subsection{Executable Regions}

Let $\{s_1,\dots,s_n\}$ be a skill library, $\mathcal{Q}$ the query space, and $\mathcal{M}$ an executor. The \emph{executable region} of skill $s_i$ is
\begin{equation}
\mathcal{R}(s_i)=\{\,q\in\mathcal{Q}\mid \mathrm{Succ}(\mathcal{M},s_i,q)=1\,\},
\end{equation}
the set of queries for which $\mathcal{M}$ succeeds when executing $s_i$. The valid routing set for query $q$ is $\mathcal{S}^{\star}(q)=\{s_i\mid q\in\mathcal{R}(s_i)\}$. If several skills can solve the query, selecting any member of $\mathcal{S}^{\star}(q)$ is correct.

Given a query distribution $P_{\mathcal{Q}}$, we measure the functional difference between two skills by
\begin{equation}
D_{\mathcal{Q}}(s_i,s_j)=\Pr_{q\sim P_{\mathcal{Q}}}
\bigl[\,q\in\mathcal{R}(s_i)\,\triangle\,\mathcal{R}(s_j)\,\bigr],
\end{equation}
where $\triangle$ denotes symmetric difference. This distance is defined by execution behavior rather than document similarity. Thus, two documents can be nearly identical even when $D_{\mathcal{Q}}(s_i,s_j)$ is large.

\subsection{Document-Imposed Retrieval Error}

A first-stage retriever computes $\hat{s}(q)=\arg\max_i r(q,x_i)$ from query $q$ and skill document $x_i$. Let $y_i(q)=\mathbf{1}[q\in\mathcal{R}(s_i)]$. After monotonically calibrating scores to $[0,1]$, define the best predictor available from the observed query and document as
\begin{equation}
g^\star(q,x_i)=\mathbb{E}\bigl[\,y_i(q)\mid q,\ X_i=x_i\,\bigr].
\end{equation}
The triangle inequality gives the diagnostic decomposition
\begin{equation}
\begin{aligned}
\bigl|\,y_i(q)-r(q,x_i)\,\bigr|
&\;\le\;
\underbrace{\bigl|\,y_i(q)-g^\star(q,x_i)\,\bigr|}_{\text{document error}}
\\
&\;\;+\;
\underbrace{\bigl|\,g^\star(q,x_i)-r(q,x_i)\,\bigr|}_{\text{retriever error}}\,.
\end{aligned}
\label{eq:decomp}
\end{equation}
The second term can be reduced by stronger encoders, fine-tuning, or reranking. The first term reflects information absent from $x_i$ and therefore remains even for an optimal reader of the document.

This limitation is explicit when $x_i=x_j$ but $\mathcal{R}(s_i)\neq\mathcal{R}(s_j)$. Any scorer receives identical observations for the two skills and must assign the same value. For any $r(q,x)\in[0,1]$,
\begin{equation}
\tfrac12\,\mathbb{E}_{q\sim P_{\mathcal{Q}}}\bigl[\,|y_i(q)-r|+|y_j(q)-r|\,\bigr]\;\ge\;\tfrac12\,D_{\mathcal{Q}}(s_i,s_j).
\label{eq:lower-bound}
\end{equation}
The bound follows from indistinguishable observations, independently of Equation~\ref{eq:decomp}. To lower it, the system must replace $x_i$ with a representation that reveals more of $\mathcal{R}(s_i)$.

\subsection{Required Discriminative Information}

For skill $s_i$, let $\mathcal{N}_i$ denote its most confusable neighbors. At the semantic level, the fields $\Tpos_i$ and $\Tneg_i$ describe query regions that should and should not select $s_i$:
\begin{equation}
\Tpos_i \subseteq \mathcal{R}(s_i),
\qquad
\Tneg_i =
\Bigl(\textstyle\bigcup_{j\in\mathcal{N}_i}\mathcal{R}(s_j)\Bigr)
\setminus \mathcal{R}(s_i).
\label{eq:tneg}
\end{equation}
The positive trigger $\Tpos_i$ describes query forms that should select $s_i$. The negative boundary $\Tneg_i$ describes superficially relevant queries that belong to neighboring skills instead. Unlike single-document expansion, constructing $\Tneg_i$ requires access to $\mathcal{N}_i$; the representation must therefore be compiled contrastively at the cluster level.

\subsection{Routing and End-to-End Success}

End-to-end success can be written as
\begin{equation}
\begin{aligned}
\mathrm{TaskSucc}(q)
&=\Pr\bigl[\hat{s}(q)\in\mathcal{S}^\star(q)\bigr]\\
&\quad\cdot\Pr\bigl[\mathcal{M}\text{ solves }q\mid \hat{s}(q)\in\mathcal{S}^\star(q)\bigr].
\end{aligned}
\label{eq:tasksucc}
\end{equation}
We use the product form in Equation~\ref{eq:tasksucc} as an attribution sketch rather than a strict identity, since an executor may occasionally succeed with a suboptimal skill. It is enough for our ablation because both arms share the same execution protocol and differ only in whether the routing card exposes $\Tneg$.

\subsection{Observed Failure Modes}
\label{sec:failure-modes}

The formulation predicts three recurring failures. \emph{Language-form mismatch} occurs when the raw document states a general rule but omits the narrative form of relevant queries; $\Tpos$ and $B$ add this positive coverage. \emph{Same-neighborhood contention} occurs when the gold skill reaches the candidate pool but loses to a functionally different neighbor with similar text; $\Tneg$ makes the boundary explicit. \emph{Keyword hijacking} occurs when a shared term raises a non-executable candidate; an exclusion condition can prevent the router from selecting it.

\begin{figure}[H]
\begin{minipage}[t]{0.46\columnwidth}
\vspace{0pt}%
\centering
\includegraphics[width=\linewidth]{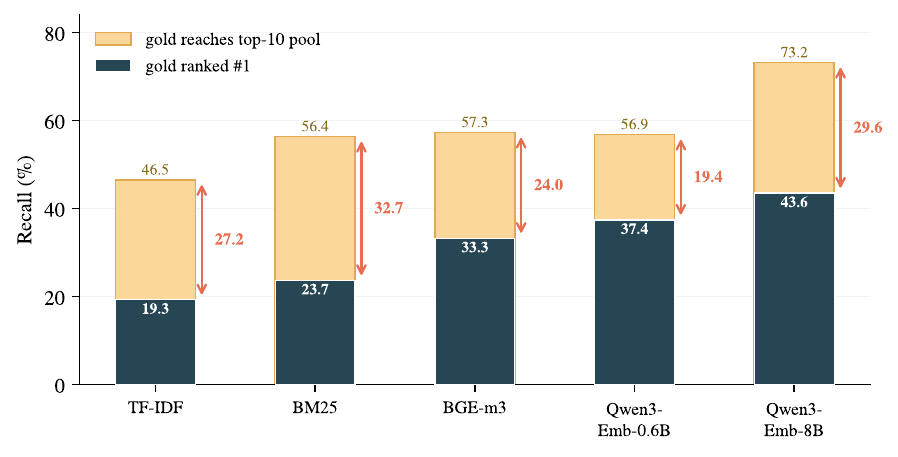}
\captionof{figure}{Top-10 pool entry versus rank-1 rate on raw documents across six datasets. The gap measures in-pool ranking error.}
\label{fig:gap}
\end{minipage}\hfill
\begin{minipage}[t]{0.50\columnwidth}
\vspace{0pt}%
Figure~\ref{fig:gap} quantifies the distinction between candidate recall and in-pool ranking using raw documents. Across six datasets, the gold skill enters the top-10 pool substantially more often than it reaches rank one. The gap remains $29.6$ points for Qwen3-Embedding-8B, showing that stronger retrieval does not eliminate candidate-level confusion. Capability Pages address the two sides separately: $\Tpos$ and $B$ target missing candidates, whereas $\Tneg$ targets errors among candidates already in the pool.
\end{minipage}
\end{figure}

\begin{figure*}[t]
\centering
\includegraphics[width=\textwidth]{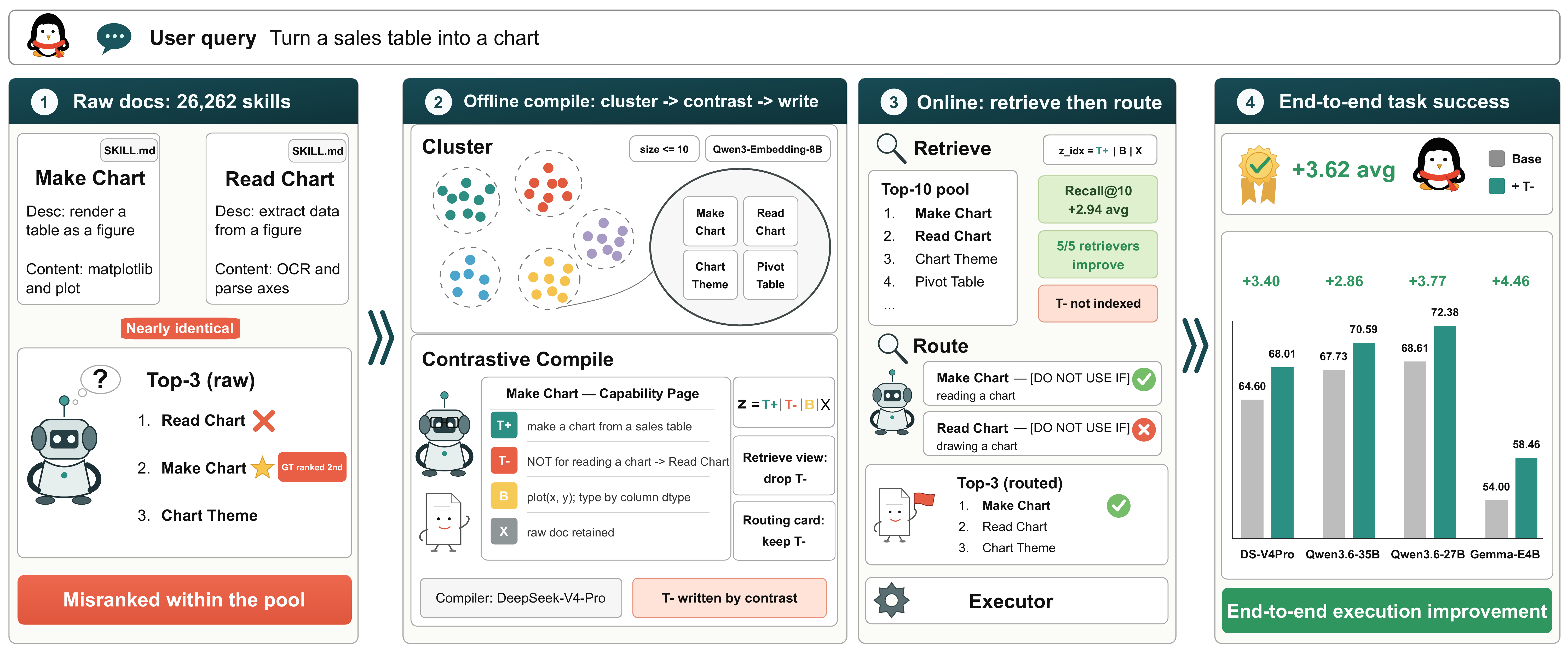}
\caption{Overview. Act I shows near-identical skill documents ranking the wrong neighbor first. Act II compiles each cluster into Capability Pages with fields $\Tpos$, $\Tneg$, and $B$. Act III uses the index view for retrieval and the $\Tneg$ card for routing. Act IV summarizes the end-to-end gain across four executors.}
\label{fig:hero}
\end{figure*}

\paragraph{A concrete same-cluster failure.}
For a MedCalc-Bench query that explicitly requests the Bazett QT correction, Qwen3-Embedding-8B ranks the Rautaharju calculator first with a similarity score of $77.79\%$ and the correct Bazett calculator second with $77.66\%$. Both documents describe QT correction from heart rate and differ mainly in the named formula: Bazett uses a square-root correction, whereas Rautaharju uses a linear correction. This near tie illustrates why topic-level similarity is insufficient and why $\Tneg$ must encode the neighbor-relative decision boundary.

\section{Method}
\label{sec:method}

Figure~\ref{fig:hero} summarizes the method. An offline compiler groups confusable skills, compares the documents within each group, and produces one Capability Page per skill. Online inference then uses separate retrieval and routing views of the page.

\subsection{Capability Page Fields}

For each skill $s_i$ with original document $x_i$, the compiler produces $z_i=(\Tpos_i,\Tneg_i,B_i,x_i)$. The fields have distinct roles:
\begin{itemize}
  \item $\Tpos_i$ describes the task type, input characteristics, and decision conditions that should route a query to $s_i$ (at most 400 characters).
  \item $\Tneg_i$ describes a confusable query form that should instead route to another skill in the same neighborhood, explicitly naming the competing capability (at most 400 characters).
  \item $B_i$ states the core formula, decision rule, or trigger pattern that distinguishes $s_i$ (at most 600 characters). It is self-contained and excludes procedures, code, and references to neighboring skills.
\end{itemize}
The original document remains part of the page so that the representation adds discriminative evidence without discarding implementation detail.

\subsection{Two-View Deployment}

The retrieval view is $z_i^{\mathrm{idx}}=(\Tpos_i,B_i,x_i)$. Sparse retrievers rebuild their index on this text, and dense retrievers encode it once offline; the query side is unchanged. We exclude $\Tneg_i$ because a similarity model may treat the neighbor description as positive content and move the skill embedding toward the very capabilities it should reject. An ablation that appends $\Tneg$ to the indexed tail confirms this design (Table~\ref{tab:index-tneg}): all three dense retrievers lose average Recall@10.

After first-stage retrieval, the router receives a compact card for each of the top-$k$ candidates. Both experimental arms include the skill name and original description. The $+\Tneg$ arm additionally exposes $\Tneg_i$ as a separate \emph{neighbor-contrast} field, while the base arm omits this field. The router compares the candidate cards and selects one skill. This design uses positive evidence to improve recall and reserves neighbor-relative boundary information for comparative selection.

\subsection{Cluster-Contrastive Compilation}

\paragraph{Clustering.}
We encode all 26{,}262 skills with Qwen3-Embedding-8B and apply two-stage clustering. KMeans with $k=200$ first partitions the library into coarse buckets. Within each bucket, agglomerative clustering with cosine-distance threshold $0.15$ forms local neighborhoods. Clusters larger than 10 are recursively split, matching both the compiler context budget and the end-to-end candidate-pool size. Skills in clusters with fewer than three members are processed individually. Dataset labels are never used. A structural sweep shows that $k\in\{100,200,300\}$ barely changes cluster statistics at a fixed cosine threshold, while the threshold $0.10$--$0.20$ controls granularity; the default $(k{=}200,\,0.15)$ lies in a stable mid-range. Neighborhoods here are used to compile offline Capability Pages, rather than to sample WRITE/SKIP annotations for retrieve--rerank training~\cite{wang2026skillisnotdoc}.

\paragraph{Contrastive generation.}
DeepSeek-V4-Pro~\citep{deepseek2026v4} reads an entire cluster and emits $\{\Tpos,\Tneg,B\}$ for every member in strict JSON. The prompt requires mutually exclusive positive triggers and negative boundaries that identify the relevant neighboring capability. Failed parses are retried; after repeated failure, the system falls back to the original document $x_i$ without generated fields. For singletons, a separate prompt infers likely confusable tasks from the document. Compilation uses only skill documents and does not access evaluation queries.

\paragraph{Quality control.}
A manual quality-control pass flags pages with unclear positive coverage or discriminative boundaries. The same cluster-level prompt regenerates only those pages, which we then audit to exclude evaluation identifiers, candidate-pool composition, and other benchmark-specific information.

\paragraph{Cost and updates.}
Each source document contains approximately $8.2$k characters on average, and the compiler emits approximately $1.1$k characters per skill. Processing the full library requires roughly $7\times10^7$ input tokens and $10^7$ output tokens. Compilation is an offline, parallelizable cost. Because each page depends only on local neighbors, adding a skill requires assigning it to a cluster and recompiling that cluster rather than the entire library.

\begin{table*}[t]
\centering\small
\begin{tabular}{llrrrrrr}
\toprule
Retriever & Indexed text & R@3 & N@3 & M@3 & R@10 & N@10 & M@10 \\
\midrule
\multirow{2}{*}{BM25}      & $x$ (raw)          & 35.28 & 33.14 & 36.56 & 56.38 & 41.02 & 40.12 \\
                           & $z^{\mathrm{idx}}$ & \textbf{39.85} & \textbf{36.86} & \textbf{39.80} & \textbf{64.01} & \textbf{45.68} & \textbf{43.62} \\
                           & \quad$\Delta$      & $+4.57$ & $+3.72$ & $+3.25$ & $+7.63$ & $+4.66$ & $+3.50$ \\
\midrule
\multirow{2}{*}{TF-IDF}    & $x$ (raw)          & 28.95 & 27.19 & 30.41 & 46.54 & 33.79 & 33.36 \\
                           & $z^{\mathrm{idx}}$ & \textbf{29.31} & \textbf{27.47} & \textbf{30.83} & \textbf{48.33} & \textbf{34.44} & \textbf{33.76} \\
                           & \quad$\Delta$      & $+0.36$ & $+0.28$ & $+0.42$ & $+1.79$ & $+0.64$ & $+0.41$ \\
\midrule
\multirow{2}{*}{BGE-M3}    & $x$ (raw)          & 45.61 & 42.60 & 45.17 & 57.32 & 47.21 & 47.18 \\
                           & $z^{\mathrm{idx}}$ & \textbf{46.42} & \textbf{43.52} & \textbf{46.17} & \textbf{58.32} & \textbf{48.26} & \textbf{48.21} \\
                           & \quad$\Delta$      & $+0.80$ & $+0.92$ & $+1.00$ & $+1.00$ & $+1.06$ & $+1.03$ \\
\midrule
\multirow{2}{*}{Qwen3-Emb-0.6B} & $x$ (raw)     & 47.49 & \textbf{45.60} & \textbf{48.85} & 56.86 & \textbf{49.38} & \textbf{50.45} \\
                           & $z^{\mathrm{idx}}$ & \textbf{47.58} & 45.33 & 48.30 & \textbf{57.61} & 49.31 & 49.99 \\
                           & \quad$\Delta$      & $+0.10$ & $-0.28$ & $-0.55$ & $+0.75$ & $-0.07$ & $-0.45$ \\
\midrule
\multirow{2}{*}{Qwen3-Emb-8B}   & $x$ (raw)     & 58.11 & 55.62 & 58.71 & 73.19 & 61.57 & 61.04 \\
                           & $z^{\mathrm{idx}}$ & \textbf{59.11} & \textbf{56.25} & \textbf{59.07} & \textbf{76.72} & \textbf{63.06} & \textbf{61.72} \\
                           & \quad$\Delta$      & $+1.01$ & $+0.63$ & $+0.35$ & $+3.53$ & $+1.49$ & $+0.68$ \\
\bottomrule
\end{tabular}
\caption{First-stage retrieval, $x$ versus $z^{\mathrm{idx}}$, averaged over six datasets, in \%. R, N, and M denote Recall, nDCG, and MRR. Better of each pair in bold.}
\label{tab:ret-main}
\end{table*}

\section{Experiments}
\label{sec:exp}

We evaluate three questions: \textbf{(RQ1)} Does the index view improve first-stage candidate recall across sparse and dense retrievers? \textbf{(RQ2)} Does the negative boundary improve end-to-end selection when the candidate pool is fixed? \textbf{(RQ3)} Does the representation transfer across language and distribution shifts?

\subsection{Experimental Setup}

\paragraph{Benchmarks.}
SRA-Bench contains 26{,}262 skills and 5{,}400 questions from TheoremQA, LogicBench, CHAMP, MedCalc-Bench, BigCodeBench, and ToolQA. SSL-SkillDiscovery provides a cross-distribution evaluation with 6{,}184 skills and 431 intent-level queries, of which $96.5\%$ are Chinese.

\paragraph{First-stage retrieval.}
We evaluate BM25~\citep{robertson2009bm25}, TF-IDF, BGE-M3~\citep{chen2024bgem3}, Qwen3-Embedding-0.6B, and Qwen3-Embedding-8B~\citep{zhang2025qwen3emb}. The raw condition indexes $x$; the proposed condition indexes $z^{\mathrm{idx}}$. The representation uses up to 4096 tokens from the name, description, and content, followed by up to 4096 tokens from $B+\Tpos$. The primary metric is Recall@10 because the router consumes a fixed pool of 10 candidates. We also report Recall, nDCG, and MRR at cutoffs 3 and 10.

\paragraph{End-to-end routing.}
The router selects one skill from the top 10 at temperature $0$. The base and $+\Tneg$ arms use identical candidate pools, prompts, and execution protocols; they differ only in whether the routing card exposes the neighbor-contrast field $\Tneg$. We test DeepSeek-V4-Pro, Qwen3.6-35B-A3B, Qwen3.6-27B-FP8~\citep{yang2025qwen3}, and Gemma-E4B~\citep{gemma2026}. Most datasets use a direct execution engine, while ToolQA uses ReAct~\citep{yao2023react}; each dataset is scored with its official evaluator.

\paragraph{Evaluation and uncertainty.}
End-to-end solving uses temperature $0.7$ and five repeated rounds per executor--dataset cell; both routing arms share the same sampling protocol. We report the mean across rounds and estimate uncertainty from the run-level standard error. The standard deviation of the per-cell gain is approximately $0.8$ points, and each executor's six-dataset mean varies by at most $0.4$ points across rounds. The confidence intervals in Table~\ref{tab:e2e} use two standard errors. Using the Student's $t$ critical value with four degrees of freedom instead widens each interval by less than $0.3$ points and leaves every executor-level lower bound above zero.

\subsection{First-Stage Retrieval}

Table~\ref{tab:ret-main} answers RQ1. Replacing $x$ with $z^{\mathrm{idx}}$ improves Recall@10 for every retriever, from $+0.75$ points for Qwen3-Embedding-0.6B to $+7.63$ points for BM25. The mean gain is $2.94$ points. Improvements extend beyond recall for four of the five models; Qwen3-Embedding-0.6B shows small decreases in nDCG@10 and MRR@10 despite higher Recall@10, indicating that the added text expands candidate coverage but can perturb top-rank ordering. This trade-off motivates handling exclusion evidence in the router rather than the vector index.

\noindent
\begin{minipage}[t]{0.48\columnwidth}
\vspace{0pt}%
\centering\footnotesize
\setlength{\tabcolsep}{3pt}
\begin{tabular}{@{}lrr@{}}
\toprule
Retriever & $\Delta$ R@10 & $\Delta$ nDCG@10 \\
\midrule
BM25 & $+0.13$ & $-0.25$ \\
TF-IDF & $+0.58$ & $+0.28$ \\
BGE-M3 & $\mathbf{-0.68}$ & $-0.22$ \\
Qwen3-Emb-0.6B & $\mathbf{-0.98}$ & $-0.71$ \\
Qwen3-Emb-8B & $\mathbf{-1.18}$ & $-0.15$ \\
\bottomrule
\end{tabular}
\captionof{table}{Six-dataset mean change when appending $\Tneg$ to $B+\Tpos$ versus the default index view.}
\label{tab:index-tneg}
\end{minipage}\hfill
\begin{minipage}[t]{0.50\columnwidth}
\vspace{0pt}%
\noindent\textbf{Index-with-$\Tneg$ ablation.}
To test the two-view claim, we append $\Tneg$ to the indexed tail ($B+\Tpos+\Tneg$) while keeping queries and retrievers fixed. Table~\ref{tab:index-tneg} reports six-dataset mean deltas relative to the default index view. All three dense retrievers lose Recall@10, while sparse effects are mixed and small. The result supports keeping $\Tneg$ out of the similarity index and exposing it only on the routing card.
\end{minipage}

\noindent
\begin{minipage}[t]{0.46\columnwidth}
\vspace{0pt}%
\centering\footnotesize
\setlength{\tabcolsep}{3pt}
\begin{tabular}{@{}lrr@{}}
\toprule
Indexed text & BM25 & Qwen3-Emb-0.6B \\
\midrule
raw & 56.38 & 56.86 \\
$+$doc2query ($N{=}5$) & 60.24 & 56.35 \\
$z^{\mathrm{idx}}$ (ours) & \textbf{64.01} & \textbf{57.61} \\
\bottomrule
\end{tabular}
\captionof{table}{Doc2query baseline, six-dataset average Recall@10 (\%).}
\label{tab:doc2query}
\end{minipage}\hfill
\begin{minipage}[t]{0.50\columnwidth}
\vspace{0pt}%
\noindent\textbf{Comparison to doc2query-style expansion.}
To test whether the index-view gains are explained by single-document query expansion alone, we compare three indexed texts under the same first-stage protocol: raw documents; raw documents appended with five LLM-generated pseudo-queries (doc2query-style, with no neighbor contrast
\end{minipage}
\noindent and no exclusion language); and our Capability Page index view $z^{\mathrm{idx}}$ ($B+\Tpos$). Table~\ref{tab:doc2query} summarizes six-dataset average Recall@10. On BM25, the averages are $56.38$ (raw), $60.24$ (doc2query, $+3.86$), and $64.01$ ($z^{\mathrm{idx}}$, $+7.63$ over raw and $+3.77$ over doc2query). On Qwen3-Embedding-0.6B, the corresponding averages are $56.86$, $56.35$ ($-0.51$ vs.\ raw), and $57.61$ ($+0.75$ over raw, $+1.26$ over doc2query). Pseudo-query expansion therefore contributes little, while cluster-compiled $B+\Tpos$ remains the stronger index representation. This is consistent with our formulation: $\Tneg$ still cannot be produced by single-document expansion and is reserved for the routing view.

\noindent
\begin{minipage}[t]{0.48\columnwidth}
\vspace{0pt}%
\centering
\includegraphics[trim=24pt 0 0 0,clip,width=\linewidth]{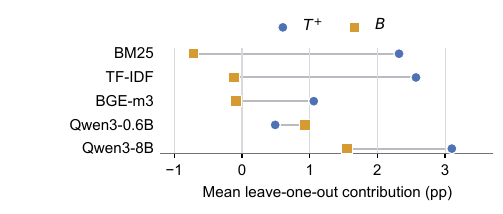}
\captionof{figure}{Mean leave-one-out contributions of $\Tpos$ and $B$ to Recall@10 over six datasets. Positive values indicate that removing the field lowers full-page recall. The field effects vary by retriever, with $\Tpos$ providing the most consistent gains.}
\label{fig:field-contrib}
\end{minipage}\hfill
\begin{minipage}[t]{0.50\columnwidth}
\vspace{0pt}%
\noindent\textbf{Which fields improve recall?}
Figure~\ref{fig:field-contrib} summarizes the leave-one-out contribution of each indexed field. Averaged over datasets, $\Tpos$ contributes positively for all five retrievers and is strongest for Qwen3-Embedding-8B. The discriminative body $B$ contributes less to sparse retrieval but is positive for both dense Qwen retrievers. The per-dataset margins in the supplement show the complementary regimes: for Qwen3-Embedding-8B, LogicBench is driven mainly by $\Tpos$ ($+9.61$ points), whereas ToolQA benefits from both $\Tpos$ ($+7.41$) and $B$ ($+6.92$). Thus, $\Tpos$ supplies query-form coverage, while $B$ provides a compact capability fingerprint.
\end{minipage}

\vspace{8pt}
\setcounter{table}{3}
\noindent
\begin{minipage}[t]{0.48\columnwidth}
\vspace{0pt}%
\centering\footnotesize
\setlength{\tabcolsep}{3pt}%
\begin{tabular}{@{}lrrrr@{}}
\toprule
Retriever & $|H_x|$ & $\rho$ (\%) & NewHit & Net \\
\midrule
BM25            & 1689 & 96.86 & 213 & $+160$ \\
TF-IDF          & 1491 & 87.59 & 203 & $+18$ \\
BGE-M3          & 2307 & 95.02 & 181 & $+66$ \\
Qwen3-Emb-0.6B  & 2368 & 95.19 &  70 & $-44$ \\
Qwen3-Emb-8B    & 2856 & 97.02 & 130 & $+45$ \\
\bottomrule
\end{tabular}
\captionof{table}{Top-1 hit preservation over six datasets. $|H_x|$ is the top-1-hit count of each retriever on the raw documents.}
\label{tab:hit-preserve}
\end{minipage}\hfill
\begin{minipage}[t]{0.50\columnwidth}
\vspace{0pt}%
\noindent\textbf{Top-1 hit preservation.}
Aggregate gains can hide regressions on queries that were already solved. Let $H_x$ and $H_z$ denote the top-1 hit sets for the raw document and Capability Page. Table~\ref{tab:hit-preserve} reports the preservation ratio $\rho=|H_x\cap H_z|/|H_x|$, the number of new hits, and the net change. Four retrievers preserve at least $95\%$ of their original top-1 hits; TF-IDF is the exception at $87.59\%$. The net top-1 change is positive for all models except Qwen3-Embedding-0.6B. For that model, Recall@10 still increases by $0.75$ points, so many displaced gold skills remain available to the downstream router.
\end{minipage}


\subsection{End-to-End Effect of the Negative Boundary}

To isolate RQ2, we use Qwen3-Embedding-0.6B as the fixed first-stage retriever because it provides the lowest dense Recall@10 ($57.61\%$) and therefore a challenging candidate pool. The router receives the same names and descriptions in both arms; only the $+\Tneg$ arm exposes the neighbor-contrast field.


Table~\ref{tab:e2e} shows an average improvement of $3.62$ points across 24 executor--dataset combinations. Every executor improves on average, and all four executor-level $95\%$ confidence intervals in Table~\ref{tab:e2e}(b) lie above zero. Gemma-E4B shows the largest mean gain ($+4.46$). Across datasets, improvements are largest on BigCodeBench ($+7.53$), ToolQA ($+5.49$), and TheoremQA ($+4.65$), while LogicBench changes by only $+0.52$ points on average. This pattern is consistent with the role of $\Tneg$: it is most useful when several candidates share vocabulary but differ in an operational condition, such as the target API behavior, tool interface, or named formula. When the candidate descriptions already expose the decisive logical rule, the additional boundary provides less new information. Because the execution protocol and candidate pools are fixed, these differences isolate the value of exposing $\Tneg$ to the router.

\subsection{Cross-Distribution Check}
\label{sec:ssl}

Table~\ref{tab:ssl-cross} answers RQ3 on SSL-SkillDiscovery. We keep Qwen3-Embedding-0.6B fixed, perform no fine-tuning, and compile Chinese $\Tpos$ and $B$ under the same field definitions used for SRA-Bench. The description-only baseline scores $66.50\%$ MRR@50. Adding $B$ raises MRR@50 to $70.10\%$, a gain of $3.60$ points, and adding $\Tpos$ further raises it to $73.07\%$. The full page reaches $73.07\%$ MRR@50, on par with the $72.95\%$ result reported by \citet{liang2026ssl} under a different evaluation pipeline; we therefore emphasize the within-method field ablation rather than a $0.12$-point leaderboard margin. Because the encoder is unchanged and no new annotations are used, the improvement

\clearpage
\setcounter{table}{4}
\begin{table}[H]
\centering\footnotesize
\setlength{\tabcolsep}{2.5pt}
\begin{tabular}{lrrrrrrrr}
\toprule
& \multicolumn{2}{c}{DS-V4Pro} & \multicolumn{2}{c}{Qwen3.6-35B} & \multicolumn{2}{c}{Qwen3.6-27B} & \multicolumn{2}{c}{Gemma-E4B} \\
\cmidrule(lr){2-3}\cmidrule(lr){4-5}\cmidrule(lr){6-7}\cmidrule(lr){8-9}
Dataset & base & $+\Tneg$ & base & $+\Tneg$ & base & $+\Tneg$ & base & $+\Tneg$ \\
\midrule
TheoremQA     & 72.16 & 73.90 & 83.00 & 85.54 & 81.93 & 84.47 & 60.51 & 72.29 \\
LogicBench    & 63.95 & 64.34 & 67.37 & 66.71 & 70.13 & 70.26 & 46.32 & 48.55 \\
CHAMP         & 76.23 & 81.61 & 82.96 & 83.86 & 82.96 & 84.75 & 78.92 & 79.82 \\
MedCalc-Bench & 89.18 & 89.82 & 89.09 & 89.82 & 90.55 & 91.18 & 72.82 & 76.00 \\
BigCodeBench  & 43.51 & 53.60 & 46.14 & 54.74 & 45.53 & 56.49 & 38.95 & 39.42 \\
ToolQA        & 42.59 & 44.76 & 37.83 & 42.87 & 40.56 & 47.13 & 26.50 & 34.69 \\
\midrule
\textbf{AVG}  & 64.60 & \textbf{68.01} & 67.73 & \textbf{70.59} & 68.61 & \textbf{72.38} & 54.00 & \textbf{58.46} \\
$\Delta$ (points) & \multicolumn{2}{c}{$\mathbf{+3.40}$} & \multicolumn{2}{c}{$\mathbf{+2.86}$} & \multicolumn{2}{c}{$\mathbf{+3.77}$} & \multicolumn{2}{c}{$\mathbf{+4.46}$} \\
\bottomrule
\end{tabular}

\textbf{(a) End-to-end task success (\%).}\\[3pt]
\begin{tabular}{lrrrrrrrc}
\toprule
& TheoremQA & LogicBench & CHAMP & MedCalc & BigCode & ToolQA & \textbf{AVG} & \textbf{95\% CI} \\
\midrule
DS-V4Pro     & $+1.74$ & $+0.39$ & $+5.38$ & $+0.64$ & $+10.09$ & $+2.17$ & $\mathbf{+3.40}$ & $[+2.69,+4.11]$ \\
Qwen3.6-35B  & $+2.54$ & $-0.66$ & $+0.90$ & $+0.73$ & $+8.60$ & $+5.04$ & $\mathbf{+2.86}$ & $[+2.29,+3.43]$ \\
Qwen3.6-27B  & $+2.54$ & $+0.13$ & $+1.79$ & $+0.63$ & $+10.96$ & $+6.57$ & $\mathbf{+3.77}$ & $[+3.15,+4.39]$ \\
Gemma-E4B    & $+11.78$& $+2.23$ & $+0.90$ & $+3.18$ & $+0.47$ & $+8.19$ & $\mathbf{+4.46}$ & $[+3.84,+5.08]$ \\
\midrule
\textbf{Col. mean} & $+4.65$ & $+0.52$ & $+2.24$ & $+1.29$ & $+7.53$ & $+5.49$ & $\mathbf{+3.62}$ & $[+3.30,+3.94]$ \\
\bottomrule
\end{tabular}

\textbf{(b) Per-cell gain $\Delta=\mathrm{Acc}(+\Tneg)-\mathrm{Acc}(\mathrm{base})$ (points).}\\[3pt]
\caption{End-to-end effect of the negative boundary. The $+\Tneg$ card improves mean task success by $3.62$ points across 24 executor--dataset combinations.}
\label{tab:e2e}
\end{table}
\setcounter{table}{5}

\noindent is attributable to the indexed representation rather than model adaptation. The incremental gains also mirror the SRA-Bench ablation: $B$ supplies a concise capability fingerprint, while $\Tpos$ aligns the document with user-facing query forms. This agreement across English and predominantly Chinese collections supports the portability of the page schema.

\noindent
\begin{minipage}[t]{0.48\columnwidth}
\vspace{0pt}%
\centering\small
\resizebox{\linewidth}{!}{%
\begin{tabular}{llr}
\toprule
Source & Indexed text & MRR@50 (\%) \\
\midrule
--- & Full SKILL.md & 64.50 \\
SSL & Desc+SSL-Rich & 72.95 \\
Ours & Description only & 66.50 \\
Ours & $+B$ only & 70.10 \\
Ours & $B+\Tpos$ (full) & \textbf{73.07} \\
\bottomrule
\end{tabular}
}
\captionof{table}{SSL-SkillDiscovery, MRR@50 in \%. The encoder is held fixed; our rows vary only the indexed page fields.}
\label{tab:ssl-cross}
\end{minipage}\hfill
\begin{minipage}[t]{0.50\columnwidth}
\vspace{0pt}%
\noindent\textbf{Cross-distribution result.}
With the encoder held fixed, $B$ raises MRR@50 by $3.60$ points over description only, and adding $\Tpos$ reaches $73.07\%$. The $72.95\%$ SSL result of \citet{liang2026ssl} uses a different evaluation pipeline, so the within-method field comparison is the appropriate transfer test. The pattern matches SRA-Bench: $B$ supplies a concise capability fingerprint, while $\Tpos$ aligns the page with user-facing query forms.
\end{minipage}

\section{Discussion}
\paragraph{Division of labor.}
The results align with the two-view design: $\Tpos$ and $B$ improve candidate coverage, whereas $\Tneg$ helps distinguish retrieved neighbors. The field ablation shows that their relative value varies by dataset, and the index-with-$\Tneg$ ablation shows why exclusion text is better reserved for the router than embedded as positive index content.

\paragraph{Scope.}
Capability Pages address the document term in Equation~\ref{eq:decomp}; stronger encoders and rerankers address the retriever term. The method therefore complements, rather than replaces, model-side improvements. Its main operational trade-off is an offline compilation and maintenance process plus longer routing cards.

\section{Limitations}
Two caveats are operational rather than fundamental. The routing benefit of $\Tneg$ applies once the correct skill reaches the top~$k$, so Capability Pages complement stronger recall rather than replace it. The clustering sweep also reports structure only, so the default setting is a stable choice rather than a tuned optimum.


\section{Conclusion}


Skill retrieval can fail when functionally different capabilities are represented by nearly indistinguishable documents. We formalized this failure through executable regions and document-imposed retrieval error, then introduced Capability Pages to expose positive triggers, discriminative rules, and neighbor-relative boundaries. Across SRA-Bench, the index view improves Recall@10 for five retrievers, and field ablations show complementary roles for $\Tpos$ and $B$. Exposing $\Tneg$ improves end-to-end success for all four executors, and the representation transfers to Chinese SSL-SkillDiscovery using a fixed encoder. Together, these results show that improving indexed and routed text can complement model-side improvements. We will release code and data upon publication.

\bibliographystyle{arxivstyle}
\bibliography{references}

\appendix
\renewenvironment{quote}
  {\par\vspace{2pt}\noindent\begingroup\small}
  {\par\vspace{2pt}\endgroup\noindent\ignorespacesafterend}
\renewenvironment{quote}
  {\par\vspace{2pt}\noindent\begingroup\small}
  {\par\vspace{2pt}\endgroup\noindent\ignorespacesafterend}
\section{Proof of the Document-Imposed Lower Bound}
\label{app:proof}

We give the calculation behind Equation~(5) in the main paper and comment on why the same construction has no natural analogue in generic document retrieval.

\paragraph{Setup.} Fix two skills $s_i,s_j$ with identical documents $x_i=x_j=:x$ and different executable regions $\mathcal{R}(s_i)\neq\mathcal{R}(s_j)$. Let $r(q,x)\in[0,1]$ be any scoring function (after monotone calibration) built from $(q,x)$. Because $x$ is the same for both skills, the scores $r(q,x_i)$ and $r(q,x_j)$ must coincide for every $q$; write their common value as $a=a(q)$.

\paragraph{Per-query bound.} On the symmetric-difference set $\mathcal{R}(s_i)\triangle\mathcal{R}(s_j)$, the labels satisfy $y_i(q)+y_j(q)=1$, so exactly one of them is $1$ and the other $0$. For any $a\in[0,1]$,
\[
|y_i(q)-a|+|y_j(q)-a| \;=\; |1-a|+|0-a| \;=\; (1-a)+a \;=\; 1.
\]
Hence on the symmetric-difference set the pairwise error is at least $1$ per query, so the per-skill average is at least $1/2$. On the complement, $y_i(q)=y_j(q)$ and the pairwise sum is $\ge 0$.

\paragraph{Averaging over $P_{\mathcal{Q}}$.} Taking expectation over $q\sim P_{\mathcal{Q}}$ and dividing by $2$ gives
\[
\tfrac12\mathbb{E}\bigl[\,|y_i-r|+|y_j-r|\,\bigr]\;\ge\;\tfrac12\Pr[q\in\mathcal{R}(s_i)\triangle\mathcal{R}(s_j)]\;=\;\tfrac12 D_{\mathcal{Q}}(s_i,s_j),
\]
which is Equation~(5) in the main paper. The bound holds for every $r$ that reads only $(q,x)$, so no encoder, fine-tuning recipe, or reranker on top of the raw document can push the average error below it. The only way to remove the bound is to change the observation, i.e., replace $x$ by a text $z_i\neq z_j$ that separates the two executable regions---the role of the Capability Page.

\paragraph{Why generic document retrieval lacks this bound.} In ad-hoc document retrieval a single query can be relevant to several documents, so labels $y_i(q)$ and $y_j(q)$ are not constrained to sum to $1$ and the step above fails. Skill routing is different: it selects one executable capability, and $y_i(q)=\mathbf{1}[q\in\mathcal{R}(s_i)]$ has an objective execution test, which is what makes the negative boundary $\Tneg$ well defined rather than a hand-crafted negative sample.

\section{Evaluation Details}
\label{app:eval}

\paragraph{Per-dataset evaluators.} The six datasets follow SRA-Bench's official scoring, all automatic at the answer level. TheoremQA and MedCalc-Bench match numbers or expressions after normalization, with a numerical tolerance; CHAMP extracts the final answer and matches exactly; LogicBench is a binary True/False match; BigCodeBench runs the generated code in a sandbox against unit tests and counts a full pass as correct; ToolQA matches the final answer string. Each cell reports $\mathrm{Acc}=\#\{\text{correct}\}/N$, where $N$ is the dataset size.

\paragraph{Execution engine.} ToolQA uses a ReAct engine; the other datasets use direct generation. For the end-to-end search-and-index setting, the indexed text is uniformly the retrieval view $z^{\mathrm{idx}}$, capped at 32768 characters and 4096 tokens; dense retrievers use their own official pooling, with last-token pooling for Qwen3-Embedding. The end-to-end candidate set is the top~10.

\paragraph{Per-dataset first-stage Recall@10.} Table~\ref{tab:ext-cmp} reports Recall@10 per dataset for all five retrievers when the index uses the Capability Page view $z^{\mathrm{idx}}$. The six-dataset average in the last column matches the $z^{\mathrm{idx}}$ column of Table~1 in the main paper.

\begin{table}[ht]
\centering\small
\setlength{\tabcolsep}{4.5pt}
\begin{tabular}{lrrrrrrr}
\toprule
Retriever & TheoQA & CHAMP & LogicB & MedCalc & BigCode & ToolQA & AVG \\
\midrule
BM25            & 83.67 & 39.09 & 45.26 & 72.82 & 66.17 & \textbf{77.06} & 64.01 \\
TF-IDF          & 69.61 & 26.87 & 17.24 & 72.18 & 63.35 & 40.70 & 48.33 \\
BGE-M3          & 86.88 & 35.91 & 18.82 & 89.91 & 60.21 & 58.18 & 58.32 \\
Qwen3-Emb-0.6B  & 93.04 & 58.96 & 28.55 & \textbf{100.00} & 51.12 & 13.99 & 57.61 \\
Qwen3-Emb-8B    & \textbf{97.72} & \textbf{69.86} & \textbf{68.29} & \textbf{100.00} & \textbf{82.56} & 41.89 & \textbf{76.72} \\
\bottomrule
\end{tabular}
\caption{Per-dataset Recall@10 with the retriever indexing $z^{\mathrm{idx}}$, in \%; best per dataset in bold.}
\label{tab:ext-cmp}
\end{table}

\paragraph{Field ablation.} Table~\ref{tab:field-ablation} reports Recall@10 for four page variants---\emph{raw}, $+\Tpos$, $+B$, and full ($z^{\mathrm{idx}}$)---per retriever and dataset, with leave-one-out margins $d(\Tpos)=\text{full}-(+B)$, $d(B)=\text{full}-(+\Tpos)$, and net gain $\Delta_{\mathrm{raw}}=\text{full}-\text{raw}$. The largest $\Delta_{\mathrm{raw}}$ cells fall on narrative datasets (LogicBench, ToolQA).

\begin{table}[ht]
\centering\footnotesize
\setlength{\tabcolsep}{3.8pt}
\begin{tabular}{llrrrrrrr}
\toprule
Retriever & Dataset & raw & $+\Tpos$ & $+B$ & full & $d(\Tpos)$ & $d(B)$ & $\Delta_{\mathrm{raw}}$ \\
\midrule
\multirow{6}{*}{BM25}
 & TheoremQA    & 82.33 & 84.20 & 82.06 & 83.67 & $+1.61$ & $-0.54$ & $+1.34$ \\
 & CHAMP        & 38.79 & 40.47 & 36.81 & 39.09 & $+2.28$ & $-1.38$ & $+0.30$ \\
 & LogicBench   & 38.55 & 48.55 & 35.53 & 45.26 & $+9.74$ & $-3.29$ & $+6.71$ \\
 & MedCalc      & 73.55 & 73.55 & 72.64 & 72.82 & $+0.18$ & $-0.73$ & $-0.73$ \\
 & BigCodeBench & 63.17 & 67.49 & 63.02 & 66.17 & $+3.14$ & $-1.32$ & $+3.00$ \\
 & ToolQA       & 77.69 & 74.13 & 80.07 & 77.06 & $-3.01$ & $+2.94$ & $-0.63$ \\
\midrule
\multirow{6}{*}{TF-IDF}
 & TheoremQA    & 68.54 & 70.68 & 67.20 & 69.61 & $+2.41$ & $-1.07$ & $+1.07$ \\
 & CHAMP        & 25.75 & 27.24 & 25.22 & 26.87 & $+1.64$ & $-0.37$ & $+1.12$ \\
 & LogicBench   & 18.55 & 20.79 & 15.13 & 17.24 & $+2.11$ & $-3.55$ & $-1.31$ \\
 & MedCalc      & 71.64 & 72.91 & 70.18 & 72.18 & $+2.00$ & $-0.73$ & $+0.54$ \\
 & BigCodeBench & 60.53 & 62.83 & 61.61 & 63.35 & $+1.74$ & $+0.52$ & $+2.82$ \\
 & ToolQA       & 34.97 & 36.22 & 35.17 & 40.70 & $+5.52$ & $+4.48$ & $+5.73$ \\
\midrule
\multirow{6}{*}{BGE-M3}
 & TheoremQA    & 86.48 & 87.28 & 86.35 & 86.88 & $+0.54$ & $-0.40$ & $+0.40$ \\
 & CHAMP        & 38.45 & 38.83 & 36.55 & 35.91 & $-0.64$ & $-2.91$ & $-2.54$ \\
 & LogicBench   & 17.24 & 19.74 & 14.87 & 18.82 & $+3.95$ & $-0.92$ & $+1.58$ \\
 & MedCalc      & 89.82 & 90.18 & 89.82 & 89.91 & $+0.09$ & $-0.27$ & $+0.09$ \\
 & BigCodeBench & 55.99 & 58.30 & 59.70 & 60.21 & $+0.52$ & $+1.91$ & $+4.22$ \\
 & ToolQA       & 57.48 & 56.15 & 56.29 & 58.18 & $+1.89$ & $+2.03$ & $+0.70$ \\
\midrule
\multirow{6}{*}{Qwen3-Emb-0.6B}
 & TheoremQA    & 91.70 & 92.50 & 92.50 & 93.04 & $+0.54$ & $+0.54$ & $+1.34$ \\
 & CHAMP        & 59.32 & 58.24 & 59.19 & 58.96 & $-0.22$ & $+0.72$ & $-0.36$ \\
 & LogicBench   & 24.47 & 25.39 & 26.32 & 28.55 & $+2.24$ & $+3.16$ & $+4.08$ \\
 & MedCalc      & 100.00 & 100.00 & 100.00 & 100.00 & $0.00$ & $0.00$ & $0.00$ \\
 & BigCodeBench & 51.14 & 49.95 & 50.70 & 51.12 & $+0.42$ & $+1.17$ & $-0.02$ \\
 & ToolQA       & 13.99 & 13.99 & 13.99 & 13.99 & $0.00$ & $0.00$ & $0.00$ \\
\midrule
\multirow{6}{*}{Qwen3-Emb-8B}
 & TheoremQA    & 96.79 & 97.59 & 96.92 & 97.72 & $+0.80$ & $+0.13$ & $+0.93$ \\
 & CHAMP        & 71.02 & 69.78 & 70.63 & 69.86 & $-0.77$ & $+0.07$ & $-1.16$ \\
 & LogicBench   & 56.32 & 65.39 & 58.68 & 68.29 & $+9.61$ & $+2.89$ & $+11.97$ \\
 & MedCalc      & 100.00 & 100.00 & 100.00 & 100.00 & $0.00$ & $0.00$ & $0.00$ \\
 & BigCodeBench & 79.94 & 83.27 & 81.04 & 82.56 & $+1.52$ & $-0.71$ & $+2.62$ \\
 & ToolQA       & 34.62 & 34.97 & 34.48 & 41.89 & $+7.41$ & $+6.92$ & $+7.27$ \\
\bottomrule
\end{tabular}
\caption{Field ablation, per-dataset Recall@10, in \%, over four page variants: raw, $+\Tpos$, $+B$, full ($z^{\mathrm{idx}}$).}
\label{tab:field-ablation}
\end{table}

\section{Full Compilation Prompts}
\label{app:prompt}

This section gives the two system prompts used at compilation, verbatim. A skill in a cluster with at least two members uses the \emph{intra-cluster contrastive} version B1, with the whole cluster as the contrast set; a skill with no semantic neighbor (a singleton) uses the \emph{singleton} version B2, where the LLM infers likely confusable tasks from the skill itself. The two share a strict JSON output format and the field length caps $\Tpos\le 400$, $\Tneg\le 400$, and $B\le 600$ characters, and differ only in the reference source for $\Tneg$. The user prompt assembles the cluster text under \code{[CLUSTER]} and \code{[FOCAL]}, each entry with skill\_id, name, description, and content, and is omitted here.

\paragraph{B1, intra-cluster contrastive, system prompt verbatim.}
\begin{quote}\small
You are compiling a CAPABILITY CARD for one skill in a large skill library. The card is the canonical retrieval and routing key for this skill, so it must be precise, conservative, and clearly distinguishable from sibling skills inside the same cluster.

\textbf{\# Reasoning policy.} Think step by step in the \texttt{thinking} channel before producing the final answer: (1) read every SIBLING and the FOCAL skill carefully---name, description, content; (2) identify what the FOCAL skill does, its input, its output, and the SINGLE canonical formula / decision rule / trigger pattern; (3) identify how it differs from EACH sibling concretely---do not write a generic ``different from related skills''; name the contrast; (4) draft the three fields, then revise once for accuracy and length; (5) only then emit the final JSON. The visible answer must be a SINGLE strict JSON object with no markdown, no code fences, and no prose.

\textbf{\# Inputs.} \texttt{[CLUSTER]}: $k$ semantically close skills, $2\le k\le 10$, each with skill\_id, name, description, content, where content may be truncated to fit the token budget; treat the body as authoritative and never invent details from a missing tail. \texttt{[FOCAL]}: the skill\_id to compile this turn; it is also one entry in \texttt{[CLUSTER]}, the others serving as its SIBLINGS, the contrast set.

\textbf{\# Output schema, STRICT.} Return exactly ONE JSON object, no markdown, code fences, or prose. Required keys, all mandatory and non-empty: \texttt{\{"T\_pos": ..., "T\_neg": ..., "B": ...\}}.

\textbf{\# Field specifications.} \emph{T\_pos}, when this skill SHOULD be invoked: one to two sentences, $\le 400$ chars; state the trigger conditions concretely, what query, which inputs, which task type; avoid vague phrasing like ``various tasks''; must be supported by the FOCAL name, description, or content. \emph{T\_neg}, when it SHOULD NOT be invoked: one to three sentences, $\le 400$ chars; must EXPLICITLY contrast the FOCAL against at least one SIBLING, referenced by skill\_id such as ``unlike web\_01234'' or by a distinctive feature; a generic ``do not use when irrelevant'' is FORBIDDEN; all cross-references to other skills live here, NOT in B. \emph{B}, the SINGLE distinctive rule identifying this skill: $\le 600$ chars, aim 300 to 500, plain text, inline math allowed, no headings, bullets, or code fences. B is a tight, dense FINGERPRINT, not a tutorial. Choose ONE template by skill type: (a) closed-form formula or theorem, the canonical equation and its conditions; (b) decision rule or clinical criterion, the rule with its thresholds and resulting decision; (c) tool, API, or dataset interface, the name and the SINGLE most distinctive trigger, which column, endpoint, file extension, or unique keyword; (d) algorithm or procedure, the name and at most two sentences on the core invariant or termination condition, without enumerating steps. HARD BANS in B: no verbatim copy of the description, no section headings, no step lists, no code snippets, no enumeration of sub-tasks, no sibling name or skill\_id.

\textbf{\# Hard rules.} (1) output STRICT JSON parseable by \texttt{json.loads}; (2) all three keys present and non-empty, no ``unsure'' escape; (3) use only information visible in \texttt{[CLUSTER]}; (4) T\_neg must contain at least one concrete sibling-level contrast; (5) B is hard-capped at 600 chars, T\_pos and T\_neg at 400; (6) English only.

\textbf{\# Self-check before emitting.} Verify: a single parseable JSON object, no code fences; three keys present and non-empty; lengths within caps; B follows exactly one template with no step lists, code, sibling names, or headings; T\_neg names a concrete sibling contrast. When it passes, emit the JSON and nothing else.
\end{quote}

\paragraph{B2, singleton, difference from B1, verbatim.} The reasoning policy, output format, length caps, $B$ templates, and hard bans of the singleton version match B1 word for word. The only substantive difference is that there are no real siblings, so the requirement on $\Tneg$ changes to:
\begin{quote}\small
This skill is a SINGLETON: it has no semantically close sibling in the library. Therefore T\_neg must instead identify likely confusable skills inferred from the FOCAL skill's own name, description, or content, such as tasks with similar inputs, similar keywords, or adjacent domains, and contrast against them. Concretely, \emph{T\_neg}: one to three sentences, $\le 400$ chars; identify one to three likely CONFUSABLE tasks, sharing keywords or surface form but differing in goal, input, or method, and contrast against them; a generic ``do not use when irrelevant'' is FORBIDDEN; all cross-references live here, NOT in B.
\end{quote}

\section{Sampling and Scoring Protocol}
\label{app:cells}

First-stage retrieval is deterministic. In the end-to-end evaluation, selection runs at temperature $0$ with a fixed seed, and solving runs at temperature $0.7$. Because solving carries sampling noise, both arms, base and $+\Tneg$, use the same repeated-sampling rule (five rounds per cell); the only difference is whether the candidate card carries $\Tneg$. We report the mean over rounds and take the standard error across runs. Run-to-run variation is small: each cell's $\Delta$ has a standard deviation of about $0.8$ points, and every executor's six-dataset mean is stable to within $0.4$ points. The $95\%$ confidence bands in the main-paper Table~3 ($\pm2$ standard errors) all lie entirely above zero, and the overall $24$-cell mean is $+3.62$ with band $[+3.30,+3.94]$. Replacing the normal factor $2$ with the exact $t_{4}$ critical value $2.78$ widens every band by under $0.3$ points and still leaves all four lower limits above zero.

\section{Capability Page Design for the SSL Check}
\label{app:ssl}

On SSL-SkillDiscovery we use Qwen3-Embedding-0.6B as a fixed encoder and apply the same page-side field design. Adding Chinese discriminative fields and positive triggers follows the same design as the Capability Page in the main text; the comparison against the SSL paper best and the $B$/$\Tpos$ ablations is reported in Table~6 in the main paper. Since most queries are Chinese, the compilation prompt keeps the same field specification as in Appendix~\ref{app:prompt} and only sets the output language to Chinese. The prompt is condensed to:
\begin{center}
\includegraphics[width=\textwidth]{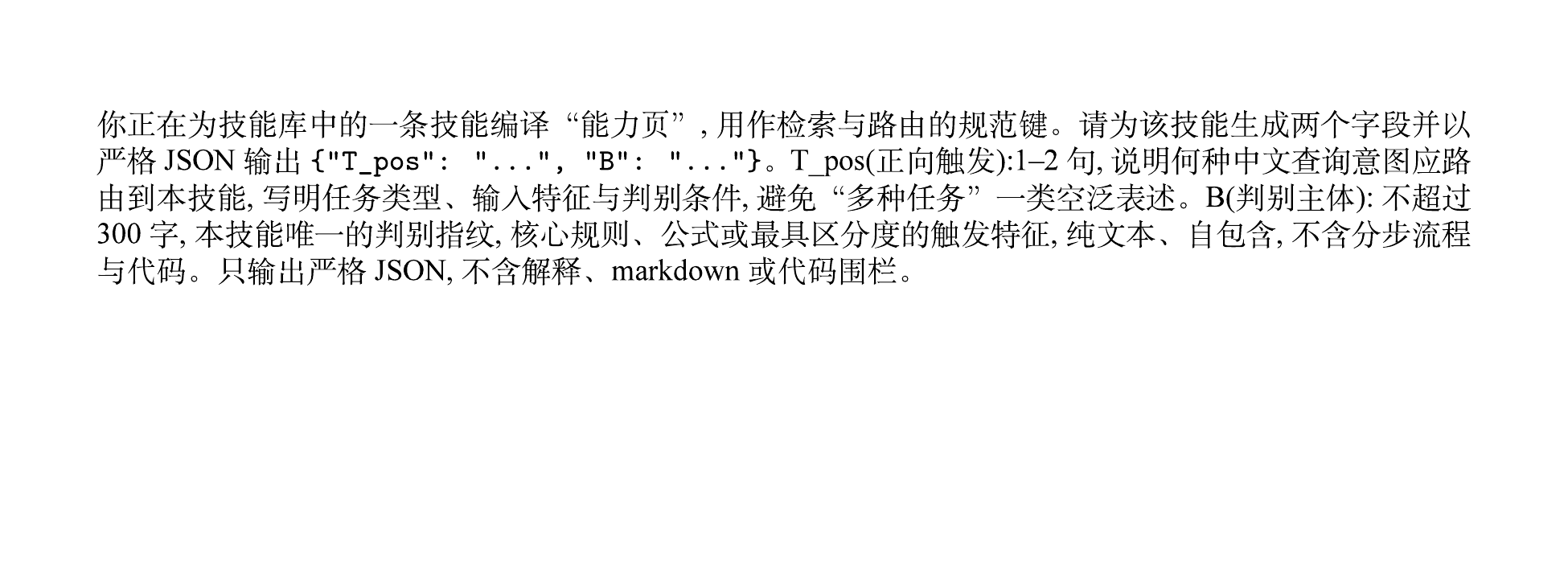}
\end{center}

\section{Raw-Document Excerpts for the Failure-Mode Examples}
\label{app:skills}

To let readers check the three failure modes in the main paper, we quote the name, description, and first content paragraph of the skills involved, transcribing math symbols for typesetting and leaving the content otherwise unchanged.

\paragraph{(a) Language-form mismatch, square identity versus subsequence limit.}
\begin{quote}\small
\textbf{champ\_005} (gold), \emph{Sum and Difference of Squares}. Description: ``Factor polynomials and bound symmetric expressions using squaring identities.'' First content paragraph: the core identity $(x\pm y)^2 = x^2 \pm 2xy + y^2$ and its rearrangement $x^2+y^2=(x+y)^2-2xy$, used to factor near-perfect-square polynomials and to bound symmetric expressions.

\textbf{champ\_020} (wrongly ranked high), \emph{Subsequence Limit Theorem}. Description: ``Use the fact that all subsequences of a convergent sequence share its limit to compute limits via simpler subsequence analysis.'' First content paragraph: if an infinite sequence converges to $L$, all its infinite subsequences converge to the same $L$; often used when a second-order recurrence is hard to analyze directly, via the odd- or even-index subsequence.
\end{quote}
The two documents share no keyword---one describes square-identity factoring and the other a subsequence limit theorem. Only because the query \code{champ\_00233} contains the word ``sequence'' do subsequence and recurrence skills such as champ\_020 rise to the front (champ\_020 at second, champ\_053 at first), while the gold champ\_005 drops out of the top~20. This is the missing positive coverage of failure mode~(a) in the main paper, to be supplied by $\Tpos$.

\paragraph{(b) Same-cluster contention, QTc skills differing only in the correction name.}
\begin{quote}\small
\textbf{medcalcbench\_020} (gold), \emph{QTc Bazett Calculator}. Description: ``Compute the corrected QT interval using the Bazett formula: QTc $=$ QT$/\sqrt{\mathrm{RR}}$, where RR $=60/$heart\_rate.'' First content paragraph: derive RR $=60/$HR in seconds, then apply Bazett QTc $=$ QT$/\sqrt{\mathrm{RR}}$.

\textbf{medcalcbench\_025} (same-cluster, wrongly first), \emph{QTc Rautaharju Calculator}. Description: ``Compute the corrected QT interval using the Rautaharju formula: QTc $=$ QT$\times(120+$HR$)/180$.'' First content paragraph: a linear correction in heart rate, QTc $=$ QT$\times(120+$HR$)/180$, with $180$ as the reference-rate baseline.
\end{quote}
Both sit in one QTc cluster separated only by the correction name, with Framingham and Hodges alongside and near-identical structure. The query \code{medcalcbench\_00180} asks explicitly for the Bazett formula, yet the gold medcalcbench\_020 comes second at $0.7766$ against medcalcbench\_025 at $0.7779$---the same-cluster contention of failure mode~(b) in the main paper, whose separating cue must come from $\Tneg$.

\paragraph{(c) Keyword hijacking, hypothetical syllogism versus self-help.}
\begin{quote}\small
\textbf{logicbench\_003} (gold), \emph{Hypothetical Syllogism}. Description: ``Chaining conditionals: if P implies Q and Q implies R, then P implies R.'' First content paragraph: the valid form $P\to Q,\ Q\to R \vdash P\to R$, extendable to a longer chain $P\to Q\to R\to S$.

\textbf{web\_08562} (wrongly first), \emph{debog-yourself}. Description: ``Help users identify and escape psychological traps. Use when the user feels stuck, unable to progress, facing a deadlock, or experiencing decision paralysis.'' First content paragraph: a self-help framework for feeling stuck, stalled, or torn.
\end{quote}
The two overlap on surface words such as ``stuck'' and ``stressed'' but have entirely different executable regions. The query \code{logicbench\_00429} is lifted by these common words so that web\_08562 goes first and the gold logicbench\_003 falls to eighth---the keyword hijacking of failure mode~(c) in the main paper, which $\Tneg$ must name and exclude.

\section{Raw Document versus Capability Page}
\label{app:capability-examples}

To show what compilation adds on top of the raw document, we place the raw description beside the compiled $\Tpos$, $\Tneg$, and $B$ for four skills, transcribed for typesetting. The first three match failure modes (a), (b), and (c) in the main paper, and the fourth shows the $B$ fingerprint of a tool skill. The $\Tpos$ translates an abstract solution rule into a problem-level query shape, the $\Tneg$ names same-cluster neighbors one by one with their skill\_id, and the $B$ compresses a unique fingerprint. None of the three can be recovered from the raw description alone.

\paragraph{(a) champ\_005, $\Tpos$ adds positive coverage.}
\begin{quote}\small
\textbf{Raw Description:} ``Factor polynomials and bound symmetric expressions using squaring identities.''

\textbf{$\Tpos$ (compiled):} Invoke when factoring a polynomial that is close to a perfect square and can be turned into a difference of squares by adding or subtracting a term, or when proving non-negativity of a symmetric expression in squared variables, or when building sums of squares via identities like $2(a^2+b^2)=(a+b)^2+(a-b)^2$.

\textbf{$\Tneg$ (compiled):} Do not use for standard factoring of a perfect-square trinomial, or for solving quadratics by completing the square, or for one-variable optimization by completing the square.

\textbf{$B$ (compiled):} Transform a near-perfect-square polynomial into a difference of squares by adding and subtracting the missing cross term, then factor as $(S+T)(S-T)$; e.g.\ $a^4+4b^4$, Sophie Germain. For symmetric reals, $(a-b)^2+(b-c)^2+(a-c)^2 = 2\sum a^2 - 2\sum ab$ yields non-negativity bounds.
\end{quote}

\paragraph{(b) medcalcbench\_020, $\Tneg$ names the same-cluster QTc neighbors.}
\begin{quote}\small
\textbf{Raw Description:} ``Compute the corrected QT interval using the Bazett formula: QTc = QT / sqrt(RR), where RR = 60 / heart\_rate.''

\textbf{$\Tpos$ (compiled):} When the task asks for the corrected QT interval under the Bazett formula, QTc = QT / $\sqrt{\mathrm{RR}}$, RR = 60 / heart rate. Invoke for queries naming ``Bazett correction'' or otherwise specifying a square-root QTc correction.

\textbf{$\Tneg$ (compiled):} Do not use if another named QTc formula is required. For linear corrections see Framingham, medcalcbench\_021, Hodges, medcalcbench\_024, or Rautaharju, medcalcbench\_025. For a cube-root correction see Fridericia, medcalcbench\_022.

\textbf{$B$ (compiled):} Bazett formula: QTc = QT / $\sqrt{60/\mathrm{HR}}$ ms. At 60 bpm, RR = 1 s and QTc equals QT. This is the canonical square-root QT correction.
\end{quote}

\paragraph{(c) logicbench\_003, $\Tneg$ blocks keyword hijacking and nearby rules.}
\begin{quote}\small
\textbf{Raw Description:} ``Chaining conditionals: if P implies Q and Q implies R, then P implies R.''

\textbf{$\Tpos$ (compiled):} Use for queries that present two or more chained conditionals sharing a middle term, ``If A then B. If B then C.'', and ask about a derived implication, including narrative ``whenever \ldots then \ldots and then \ldots'' chains.

\textbf{$\Tneg$ (compiled):} Do not use for a single conditional with a direct premise, modus ponens or tollens, logicbench\_000 and logicbench\_001. Do not use when the query denies the antecedent, since $\neg$A $\vdash$ $\neg$C is invalid. Do not use for judging premise truth or for non-deductive reasoning.

\textbf{$B$ (compiled):} Hypothetical Syllogism, transitivity of implication: $P\to Q$, $Q\to R \vdash P\to R$. Do NOT infer $\neg$R from $\neg$P, since denying the antecedent of the chain is invalid.
\end{quote}

\paragraph{(d) bigcodebench\_025, the $B$ fingerprint of a tool skill.}
\begin{quote}\small
\textbf{Raw Description:} ``sqlite3 connection patterns, CREATE TABLE IF NOT EXISTS, parameterized queries, pandas integration, and column ordering pitfalls.''

\textbf{$\Tpos$ (compiled):} Triggered when the task needs programmatic creation, population, or querying of a local SQLite database in Python, parameterized inserts with ? placeholders, or loading SQL results into pandas via read\_sql\_query.

\textbf{$\Tneg$ (compiled):} Do not invoke for interactive ad-hoc queries in the sqlite3 command-line shell, for ORM access such as SQLAlchemy, or for tasks relying solely on pandas.DataFrame.to\_sql.

\textbf{$B$ (compiled):} Python stdlib sqlite3: connect with sqlite3.connect, run CREATE TABLE IF NOT EXISTS and INSERT with ? parameter binding via cursor.execute, commit, and read into pandas via pd.read\_sql\_query inside the connection context, which gives explicit schema control, safe value binding, and correct column ordering.
\end{quote}

\end{document}